\documentclass[twocolumn,showpacs,preprintnumbers,amsmath,amssymb,pra]{revtex4}

\usepackage{graphicx}
\usepackage{dcolumn}
\usepackage{bm}
\usepackage{aas_macros}
\usepackage{changebar}

\newcommand{\figwidth}{8.0cm}   

\def\pd{\partial}
\def\be{\begin{equation}}
\def\ee{\end{equation}}
\def\ul{\underline}

\begin{document}


\title{
Revised Born-Oppenheimer approach and a multielectron reprojection method for inelastic collisions
}%

\author{Andrey K. Belyaev}
\affiliation{%
Department of Theoretical Physics,  Herzen University, St.~Petersburg 191186, Russia 
}%

\date{\today}

\begin{abstract}
The quantum reprojection method 
within the standard adiabatic Born-Oppenheimer approach 
is derived for multielectron collision systems. 
The method takes nonvanishing asymptotic nonadiabatic couplings into account
and distinguishes asymptotic currents in molecular state and in atomic state channels, 
leading to physically consistent  and reliable results. 
The method is demonstrated for the example of low-energy inelastic Li+Na collisions, 
for which the conventional application of the standard adiabatic Born-Oppenheimer approach fails 
and leads to paradoxes such as infinite inelastic cross sections. 
\end{abstract}

\maketitle

The majority of theoretical treatments of 
inelastic collisions involving atoms, ions, molecules, clusters, surfaces, etc,
is performed within the standard adiabatic Born-Oppenheimer (BO) approach (or simply the BO approach), 
which is described, e.g., in \cite{MaciasRiera:82physrep,BransdenMcDowell:92}. 
The approach is based on the separation of the electronic and nuclear motion.   
At first, the electronic fixed-nuclear Hamiltonian is treated and 
the electronic molecular states are determined, 
then the nuclear dynamics is studied 
using the expansion of the total wave functions in terms of electronic molecular-state wave functions. 
The BO approach gives a clear physical picture of the scattering process 
and allows one to use of the well-developed quantum-chemical methods and 
computer programs.

Although the BO approach looks straightforward, it encounters severe difficulties.
The problem was first recognized in Ref.~\cite{BatesMcCarroll:58} 
and became known as the "electron translation (ET) problem". 
In fact, the name of the "molecular-state problem" is more adequate for this case, 
because the origin of this problem is the use of molecular states 
(which is the basis of the BO approach) in scattering process treatments. 
The proposed remedies for the ET problem are essentially based on: 
(i) the inclusion of ET factors \cite{BatesMcCarroll:58,ThorsonDelos:78a} 
or common translation factors \cite{SchneidermanRussek:69,AllanHanssen:85} 
into the expansion of wave functions,
or (ii) the use of state-specific reaction coordinates 
\cite{Mittleman:69,ThorsonDelos:78b,GargaudMcCarrollValiron:87}, 
(specific) hyperspherical coordinates \cite{SolovevVinitsky:85}, 
Eckart coordinates \cite{RobertBaudon:86}, etc. 
The methods were reviewed in 
Refs.~\cite{MaciasRiera:82physrep,BransdenMcDowell:92,McCarrollCrothers:94,Errea-etc:94}. 
The remedies lead to modifications of basis functions, potentials, couplings, and dynamical equations 
 and are conceptionally rather complicated when compared with the original concept. 
It has been stated that it is "not possible to extract a meaningful scattering matrix" \cite{McCarrollCrothers:94}, 
and finally the problems have been interpreted as conceptional limitations of the entire 
BO approach \cite{BransdenMcDowell:92}. 
Nowadays, this fundamental problem still presents unresolved features, e.g., infinite scattering lengths 
in ultralow-energy collisions \cite{Wolniewicz:03pra}. 
Efforts to solve the ET problem have been  continued, 
and, in particular, the {\em one-electron} quantum 
reprojection method has been derived \cite{GMB:99pra,BEGM:01pra,Bel:09physscr}, 
which is conceptionally rather simple and uses BO molecular potentials and couplings.
This method is generalized in the present paper for a multielectron case.

In most applications, however, the ET problem is simply neglected, that is, 
(i) all asymptotic nonadiabatic couplings are cut off at a finite internuclear distance, 
and (ii) the asymptotic boundary conditions are taken in the BO (see below) or similar form,  see, 
e.g., \cite{Stenrup-etc:09pra}. 
Let us refer to this application of the standard adiabatic BO approach as the "conventional BO method", 
because the majority of cross-section calculations is carried out with this procedure, 
in contrast to the above mentioned methods which take the ET problem into account. 
When applying the conventional BO method, it is assumed that the approximations described above 
give negligible errors, at least for low collision energies, see, e.g., \cite{Stenrup-etc:09pra, Liu-etc:10pra}. 
Nevertheless, as shown in Refs.~\cite{Bel:09physscr, BVK:07os}, even at low collision energies 
the conventional BO method leads to the paradoxes such as 
nonzero nonadiabatic transition probabilities and nonzero inelastic cross sections for noninteracting 
collisional systems, e.g., n + H.

Taking as an example Li+Na collisions, the present paper shows 
that in some cases no proper cutoff can be found to get reliable results 
and that the above mentioned approximations can lead to errors which are several orders of magnitude larger than correct values 
even at low collision energies. 
Moreover, it is shown that in case of nonzero asymptotic couplings, 
which are the rule rather than the exception of the BO approach, 
the conventional BO method leads to infinitely large inelastic cross sections. 
Therefore, the coupling-cutoff procedure 
can give {\em any value} for an inelastic cross section.

For the sake of simplicity, let us treat atomic collisions in $\Sigma$ molecular states. 
Within the BO approach 
the total wave function $\Psi_{J\,M_J}({\bf r},{\bf R})$ is expanded as 
\begin{equation}
   \Psi_{J\,M_J}({\bf r,R})= Y_{J\,M_J}(\Theta,\Phi)
                    \sum_k \frac{F_k(R)}{R}\,\phi_k({\bf r},{\bf R})
                                              \, ,  \label{eq:expansion}
\end{equation}
$\phi_k({\bf r},{\bf R})$ being the electronic molecular-state wave functions, 
${\bf r}$ and ${\bf R}$ being the sets of electronic and nuclear coordinates, 
$J$, $M_J$ being the total angular momentum quantum numbers.  
This results in a system of coupled channel equations (CCE) for radial nuclear wave functions $F_k(R)$, 
see, e.g., \cite{MaciasRiera:82physrep}. 
Nonadiabatic transition probabilities are then calculated 
in the asymptotic ($R\to \infty$) region, where the conventional BO method assumes 
that an incoming/outgoing current in a {\em single atomic state} proceeds
completely into a {\em single molecular state} and vice versa. 
The asymptotic boundary conditions for the total wave function read
\begin{equation}
   \Psi_{J\,M_J}({\bf r},{\bf R}) =
   \sum_{j} K_{j}^{-1/2}\left(a_{j}^+\Psi_{j}^+
                   +a_{j}^-\Psi_{j}^-\right) 
                                                      \, , \label{eq:bound2}
\end{equation}
with the wave numbers $ K_{j}$ and the outgoing/incoming amplitudes $a_j^{\pm}$ in the atomic-state channel $j$.  
The conventional BO method 
assumes the following outgoing and incoming asymptotic ($R\to \infty$) BO wave functions 
\begin{equation}
   ^{BO}\Psi_{j}^{\pm}=\frac{\exp(\pm iK_{j}R)}{R}\, 
   Y_{JM_J}(\Theta,\Phi) \  \phi_j({\bf r},{\bf R})  
                                    \, .                \label{eq:BOwfpm}
\end{equation}
The CCE are solved numerically from zero up to an upper integration 
limit $R_0$, which is large enough to calculate transition probabilities  
based on the asymptotic wave functions (\ref{eq:BOwfpm}) \cite{MaciasRiera:82physrep}.

The use of electronic molecular states leads to the following fundamental features. 
Both radial and rotational nonadiabatic couplings, see, e.g.,  
\cite{MaciasRiera:82physrep, Dalgarno-etc:81pra, GMB:99pra, BDM:02jcp}, 
as well as the form of CCE 
\cite{GMB:99pra, BDM:02jcp} depend on the origin of the electron coordinates,
although the CCE themselves are independent on the origin choice
\cite{Zygelman-etc:92pra,  GMB:99pra, BDM:02jcp}. 
The CCE take their standard and  
simplest form in Jacobi coordinates, where 
the electrons are measured from the center of nuclear mass (CNM) (neglecting the mass-polarization term). 
The asymptotic values of the radial nonadiabatic couplings 
calculated with the electron origin at the CNM read, 
see, e.g., \cite{GMB:99pra, BEGM:01pra}, 
\be
   \langle {j}|\frac{\pd}{\pd R}|{k}\rangle_{\infty}=
   \gamma_{k}\,\frac{m}{\hbar^2}\,\left[V_{j}(\infty )-V_{k}(\infty )\right] 
   \langle{j}|d_z^{at}|{k}\rangle
                                           \, , 
                                                  \label{eq:asymptrad}
\ee
$\langle{j}|d_z^{at}|{k}\rangle$ being the atomic transition dipole moment, 
$m$ being the electron-nuclei reduced mass, $V_{j}(R)$ being an adiabatic potential. 
The scalar factors $\gamma_{k}$ depend on which nucleus an active electron 
is bound in the asymptotic region: 
\be
   \gamma_{k} = \left\{
   \gamma_A = -\frac{M_B}{M_A+M_B} \, 
                                 {\rm ,~an~electron~bound~with~}A
   \atop
   \gamma_B = +\frac{M_A}{M_A+M_B} \, 
                              {\rm ,~an~electron~bound~with~}B.
              \right. 
                                                    \label{eq:gamma} 
\ee
$A$ and $B$  label of nuclei with masses $M_{A}$ and $M_{B}$.
About rotational couplings, see \cite{BEGM:01pra}.
It is seen that some radial nonadiabatic couplings 
remain nonzero in the asymptotic region 
even for the noninteracting model system n+H \cite{Bel:09physscr,BVK:07os,BDM:02jcp}.  
It is worth to emphasize that choosing another electron origin (e.g., at one of the nuclei) 
does not help to avoid nonzero asymptotic couplings in the CCE, because 
the same nonzero values appear in the equations due to new terms in the Hamiltonian 
\cite{GMB:99pra, BDM:02jcp}.
According to the BO approach, nonzero  couplings provide 
transitions between molecular states even at $R\to\infty$.

In fact, the nonvanishing asymptotic couplings are a consenquence of a more fundamental shortcoming. 
The coordinates used to describe molecular states of the collision complex 
at small and intermediate distances are not suited
for the description of the free atoms in the asymptotic region. 
The correct asymptotic incoming/outgoing wave functions 
\cite{MaciasRiera:82physrep,BatesMcCarroll:58,McCarrollCrothers:94,GMB:99pra,BEGM:01pra} 
\begin{equation}
   \Psi_{j}^\pm=\frac{\exp(\pm iK_{j}R_j^{at})}{R_j^{at}}\, 
   Y_{JM_J}(\Theta,\Phi) \  \phi_j 
                                                      \label{eq:wfpm}
\end{equation}
are written in another set of Jacobi coordinates 
and different from the BO functions  \eqref{eq:BOwfpm}. 
The vector ${\bf R}_j^{at}$ connects the centers of mass of
the atoms, in contrast to ${\bf R}$ which connects the nuclei.

The reprojection method for multielectron collision systems consists in the following.   
The vector ${\bf R}_j^{at}$ can be written as follows
\begin{equation}
   {\bf R}_j^{at}={\bf R} + {\bf b}_j 
                                                      \, , \label{eq:Rat}
\end{equation}
where the vector ${\bf b}_j$ is equal to a sum over all electrons 
\begin{equation}
   {\bf b}_j = \sum_{\alpha} \gamma_{j}^{\alpha}\frac{m_j^{\alpha}}{M}~
             \left({\bf r}^{\alpha}-\gamma_{j}^{\alpha}{\bf R}\right) 
                                                      \, , \label{eq:vector-b}
\end{equation}  
index $\alpha$ labeling 
the electrons, ${\bf r} = \{{\bf r}^\alpha \}$. 
$m_j^{\alpha}$ is the electron-nucleus reduced mass in the channel $j$:
$m_j^{\alpha}=m_eM_A/(m_e+M_A)$, if the electron ${\alpha}$ is bound to nucleus $A$, and
$m_j^{\alpha}=m_eM_B/(m_e+M_B)$, if it is 
bound to nucleus 
$B$;
$m_e$ being the electron mass; 
$M$ being the reduced mass of the nuclei. 
The shift ${\bf b}_j$ depends on the asymptotic electron rearrangement in the channel $j$ 
and, hence, can be different for different channels. 
It is small, but it does not vanish at infinity 
and therefore should be taken into account.

A single term in the expansion \eqref{eq:expansion}, 
describing a current in the molecular state $k$, 
does not coincide with a single term 
in Eq.~\eqref{eq:bound2}, describing a current in the corresponding atomic state, 
see Eq.~\eqref{eq:wfpm}.  
An incoming/outgoing current in a {\em single atomic state} 
is distributed among {\em several molecular states} and vice versa. 
Projecting of the atomic-channel asymptotic wave functions (\ref{eq:wfpm})
on the molecular asymptotic wave functions (\ref{eq:BOwfpm}) gives 
\begin{equation}
   \Psi_{j}^\pm=\frac{\exp(\pm iK_{j}R)}{R}\, 
   Y_{JM_J}(\Theta,\Phi) \ \sum_k t^{\pm}_{kj}\, \phi_k({\bf r},{\bf R})
                                                 \, ,      \label{eq:wfpmproj}
\end{equation}
where the elements of the matrices $\underline{\underline{t}}^\pm$ 
represent the reprojection coefficients 
\begin{equation}
   t_{kj}^\pm=\left\langle k\left| \exp{\left(\pm i \, K_j \, {b_j}_z\right)} 
               \right|j \right\rangle_{\infty}
                                      \, .    \label{eq:t-matrix}
\end{equation}
At low collision energies these matrix elements can be approximately evaluated via 
corresponding atomic transition dipole moments and furthermore, taking into account 
Eq.~(\ref{eq:asymptrad}), via asymptotic values 
of the derivative couplings calculated in the Jacobi molecular coordinates 
\begin{equation}
   t_{kj}^\pm=\delta_{kj}\pm \frac {i K_j\hbar^2}{M\left(V_k(\infty)-V_j(\infty)\right)}
   \left\langle k\left|\frac{\partial}{\partial R}\right| j\right\rangle_{\infty}
                                      \, ,    \label{eq:asytpm}
\end{equation}
where all values are taken in the asymptotic region.  
So, the asymptotic couplings are responsible for the correct asymptotic wave functions 
in the ${\bf r}$, ${\bf R}$ coordinates.

A numerical solution of the CCE for a molecular channel 
at $R\to\infty$ gives a superposition of 
the BO asymptotic wave functions in atomic channels.  
Within the reprojection method, 
the CCE with nonzero asymptotic nonadiabatic couplings are integrated numerically 
from zero up to a large 
distance $R_0$ resulting in the $R$-matrix ($\ul{\ul{R}}$). 
Finally, taking into account Eq.~(\ref{eq:wfpmproj}), 
the $S$-matrix is expressed via the $R$-matrix as follows
\begin{eqnarray}
   \underline{\underline{S}} & = & (-1)^J\exp(-i\underline{\underline{K}}R_0)\, 
       \underline{\underline{K}}^{-1/2}\,
       \left(\underline{\underline{t}}^- + i\underline{\underline{R}}\,
       \underline{\underline{t}}^-\underline{\underline{K}}\right)
        \nonumber \\
&& \times       
       \left(\underline{\underline{t}}^+ - i\underline{\underline{R}}\,
       \underline{\underline{t}}^+\underline{\underline{K}}
       \right)^{-1}
       \underline{\underline{K}}^{1/2}\exp(-i\underline{\underline{K}}R_0)
                                                       \, .
                                                        \label{eq:rtos}
\end{eqnarray} 
The formula (\ref{eq:rtos}) is valid for multielectron collisions and 
turns into the formula derived previously in the limiting one-electron case \cite{GMB:99pra, BEGM:01pra}. 
The difference between the conventional BO method and the reprojection method is in the presence of 
the $t^{\pm}$-matrices in the latter instead of the unit matrix in the former, see Eq.~(\ref{eq:rtos}). 
The implementation of the reprojection method 
(also called as the $t$-matrix method)
is not more complicated than the implementation of the conventional BO method 
and does not require additional input data if Eq.~(\ref{eq:asytpm}) is used.

%
\begin{figure}
\vspace*{-2mm}
\includegraphics[angle=0, width=10.0cm]{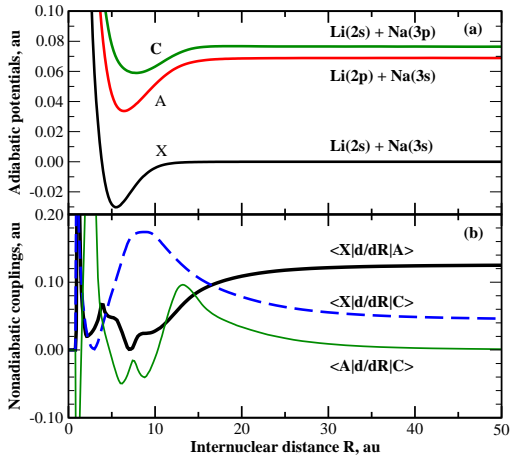}  
\caption{\label{fig:qc-data}
The adiabatic potentials (upper panel, a) for the three lowest LiNa($^1\Sigma^+$) states ($X,\, A$, $C$) 
and the radial nonadiabatic couplings (low panel, b) between these states. 
}
\end{figure}
%
Let us consider Li+Na collisions. 
The adiabatic potentials for the three lowest LiNa($^1\Sigma^+$) states  ($X,\, A$, and $C$) 
and the radial nonadiabatic couplings between them are plotted in Fig.~\ref{fig:qc-data}. 
The {\em ab initio} potentials are taken from Ref.~\cite{Petsalakis-etc:08jcp}, 
while the nonadiabatic couplings have been calculated in the present work by means of 
the MOLPRO package (the {\em ab initio} MRCI method).
Fig.~\ref{fig:qc-data} clearly shows that two nonadiabatic couplings remain nonzero 
in the asymptotic region in agreement with Eq.~(\ref{eq:asymptrad}). 
Their values are not negligible compared with the typical maximum value of $\approx 0.2$~a.u. 
Moreover, the $X$-$A$ coupling has its maximum value in the asymptotic region 
(except for the range $R<2$~a.u., which is not important for transitions). 
Thus, no proper cutoff can be found to get a reliable result.

\begin{figure}
\includegraphics[angle=0, width=\figwidth]{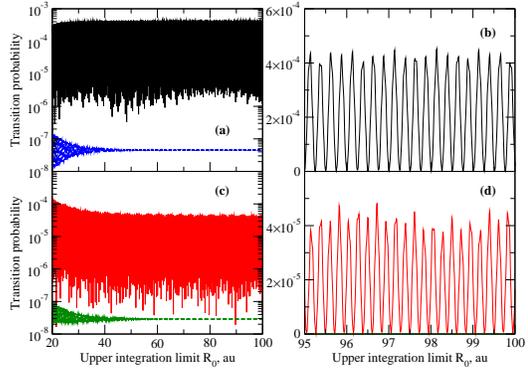}
\caption{\label{fig:P-Rstop}
The transition probabilities for ${\rm Li}(2s\to 2p) + {\rm Na}(3s)$ 
(upper panels) 
and 
${\rm Li}(2s) + {\rm Na}(3s\to 3p)$ 
(low panels) excitation calculated for 
$E=5$~eV and $J=0$ by means of 
the conventional BO method (the solid black and red lines) and by means of 
the reprojection method 
(the dashed blue and green lines) as a function of the upper integration limit $R_{0}$. 
The right panel shows the probabilities in the enlarged scale. 
}
\end{figure}

The transition probabilities  $P_{if}(J,E)$ for 
the 
${\rm Li}(2s) + {\rm Na}(3s) \to {\rm Li}(2p) + {\rm Na}(3s)$ 
and 
${\rm Li}(2s) + {\rm Na}(3s) \to {\rm Li}(2s) + {\rm Na}(3p)$ 
excitation processes 
are shown in Fig.~\ref{fig:P-Rstop} 
for  the collision energy 
$E=5$~eV and $J=0$ 
as a function of the upper intergation limit $R_0$.   
%
%
It is seen that $P_{if}(J,E)$ 
calculated by means of the conventional BO method 
oscillate between roughly zero and relatively large values 
with increasing $R_0$. 
The variations represent 
nonadiabatic transitions between molecular states at large $R$ 
and are the consequence of the nonzero asymptotic couplings. 

%
\begin{figure}
\vspace*{-1mm}
\includegraphics[angle=0, width=\figwidth]{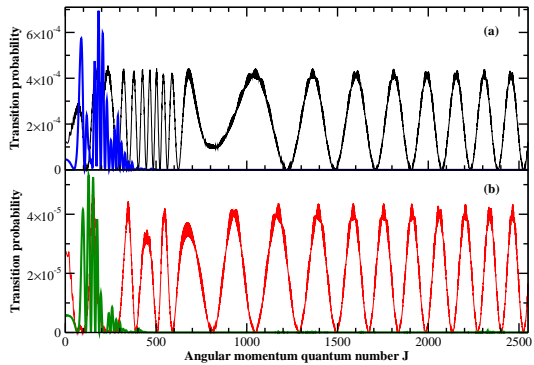}
\caption{\label{fig:PJ}
The transition probabilities for 
${\rm Li}(2s\to 2p) + {\rm Na}(3s)$ (a) 
and 
${\rm Li}(2s) + {\rm Na}(3s\to 3p)$ (b) excitation 
as a function of the total angular momentum quantum number $J$ 
for $E=5$~eV. 
The thin black and red lines show the probabilities obtained by means of the conventional BO method; 
the thick blue and green lines depict the probabilities calculated 
by the reprojection method  
and multiplied by 1000 (a) and 200 (b).  
}
\end{figure}
%

The reprojection $t$-matrix method yields transition probabilities 
which are independent on the upper integration limit, 
when it is large enough, see Fig.~\ref{fig:P-Rstop}.   
The transition probabilities are 
several orders of magnitude smaller 
those obtained by the conventional BO method and are not equal to averaged values of the latter. 
Nonadiabatic transitions between {\em molecular states} still remain 
at an arbitrary large $R$, but they do not produce transitions between 
{\em atomic states} at large distances. 
The $t$-matrices correct the $S$-matrix (\ref{eq:rtos}). 
Thus, transitions between {\em atomic states} in the asymptotic region are unphysical, 
while transitions between {\em molecular states} in the same region are physical 
and represent a fundamental feature of the BO approach.

The same  transition 
probabilities $P_{if}(J,E)$ as a function of 
$J$ are plotted in Fig.~\ref{fig:PJ} for $E=5$~eV, $R_0=500$~a.u. 
The conventional BO method gives probabilities which remain oscillating with 
increasing $J$ due to the nonzero asymptotic couplings: 
at any $J$ a centrifugal term does not prevent reaching a nonadiabatic region. 
In contrast to this, 
the transition probabilities obtained by means of the reprojection 
method are 
substantial only within a limited range of $J$, 
roughly 
up to $J\approx 500$ for $E=5$~eV.  
Note 
the probabilities calculated by the reprojection  
method and plotted in Fig.~\ref{fig:PJ} 
are multiplied by 1000 and 200.

%
\begin{figure}[t]
\includegraphics[angle=0, width=\figwidth]{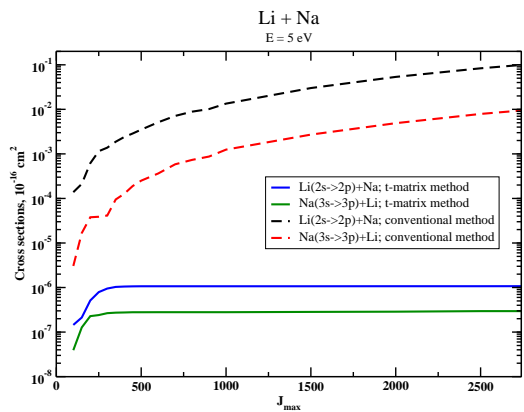}
\caption{\label{fig:sigma-Jmax}
The ${\rm Li}(2s\to 2p) + {\rm Na}(3s)$ (black and blue lines)  
and 
${\rm Li}(2s) + {\rm Na}(3s\to 3p)$ (red and green lines) 
excitation cross sections 
calculated by means of the conventional BO method (dashed lines) and 
by means of the  reprojection 
$t$-matrix method (solid lines) 
as a function of the maximum value of 
the angular momentum quantum number $J_{max}$
for $E=5$~eV.
}
\end{figure}
%
The inelastic cross sections are calculated as a sum over 
$J$ from 0 till infinity or 
a value $J_{max}$ where a convergence is reached. 
If the range of $J$ with nonzero transition probabilities is unlimited, as shown 
in Fig.~\ref{fig:PJ} for the conventional BO method results,  
convergence cannot be reached and cross sections infinitely increase  
with increasing of the upper summation limit $J_{max}$, as depicted in Fig.~\ref{fig:sigma-Jmax}. 
Thus, 
nonzero asymptotic couplings lead to infinite 
inelastic cross sections obtained with the conventional BO method. 
This conclusion holds even in case of small nonzero asymptotic couplings.

In the framework of the reprojection $t$-matrix method, the nonadiabatic transition probabilities $P_{if}(J,E)$ 
are nonzero only within a limited range of $J$ (as it must be), 
which leads to the convergence of the cross sections with increasing of 
$J_{max}$, 
see Fig.~\ref{fig:sigma-Jmax}, and finally to the finite values of the inelastic 
cross sections. 
All remedies for the ET problem are supposed to lead to the convergence. 
Thus, the conventional BO method has its limitation both in the formalism and in its applications, 
while the standard adiabatic BO approach with any ET remedy is free from such limitations.

It should be mentioned that a lack of convergence with respect to an integration limit variation
within the conventional BO method at high energies 
was noticed in Ref.~\cite{AllanHanssen:85} for another processes. 
In those papers it was found that some variation of integration limits 
results in increasing of the total cross sections up to a factor of 3 at high collision energies, 
while at low energies ($<300$~eV) "the total cross sections without common translation factors are only slightly different."
In the same papers it was found that the inclusion of the common translation factors leads to the convergence 
with respect to an integration limit variation. 
In the present work it is found that within the conventional BO method there is no convergence 
with respect to both an upper integration limit $R_0$ variation (Fig.~\ref{fig:P-Rstop}) 
and a variation of the maximum value of the angular momentum quantum number $J_{max}$ 
(Fig.~\ref{fig:sigma-Jmax}) even at low collision energies, 
while the reprojection $t$-matrix method provides the convergence with respect to variation of both $R_0$ and $J_{max}$, 
see Figs.~\ref{fig:P-Rstop} and \ref{fig:sigma-Jmax}. 
Moreover, it is shown that at low collision energies, the results of calculations with and without 
taking into account the ET (or molecular-state) problem differ not only by a few times, but by several orders of magnitude.

It has thus been demonstrated that the conventional BO method applied 
to collision processes with nonzero asymptotic nonadiabatic couplings, which are fundamental features of the BO approach, 
leads to paradoxes such as infinite inelastic cross sections even at low collision energies. 
The reprojection method takes into account nonzero asymptotic couplings, 
which are responsible for correct asymptotic wave functions,  
and distinguishes the asymptotic currents in {\em molecular} and {\em atomic}  state channels, 
providing physically consistent and reliable results.

The author thanks Prof. W. Domcke for fruitful discussions. 
The work was supported by the Russian Foundation for Basic Research (Grant No. 10-03-00807-a)
and the University of Rome "La Sapienza".

\end{document}